\documentclass[12pt]{article}

\usepackage{cite}
\usepackage{color}
\usepackage{graphicx}
\usepackage{amsmath}
\usepackage{amssymb}
\usepackage{xspace}

\makeatletter
\@addtoreset{equation}{section}

\makeatletter
\renewcommand\section{\@startsection {section}{1}{\z@}%
                                   {-3.5ex \@plus -1ex \@minus -.2ex}
                                   {2.3ex \@plus.2ex}%
                                   {\normalfont\large\bfseries}}
\renewcommand\subsection{\@startsection{subsection}{2}{\z@}%
                                     {-3.25ex\@plus -1ex \@minus -.2ex}%
                                     {1.5ex \@plus .2ex}%
                                     {\normalfont\bfseries}}

\def\baselinestretch{1.2}
\def\Reals{{\mathbb{R}}}

\newcommand{\be}{\begin{equation}}
\newcommand{\ee}{\end{equation}}
\newcommand{\beq}{\begin{eqnarray}}
\newcommand{\eeq}{\end{eqnarray}}

\newcommand{\tr}{{\rm Tr}}
\newcommand{\gone}[1]{{}}
\newcommand{\sty}[1]{{{\rm #1}}}

\def\anglephi{{\phi}} 
\def\dilaton{{\Phi}} 

\def\fullu{{u}}  
\def\rescaledU{{U}} 
\def\rescaledY{{Y}} 

\def\basephi{{\varphi}} 
\def\basetheta{{\theta}} 
\def\fiberphi{{\phi}} 

\def\anglepsi{{\psi}} 
\def\angleY{{\xi}} 
\def\Mcircle{{y_{10}}} 

\def\radialV{{V}}

\newcommand\dsBase[1]{ds_{B({#1})}} 
\def\connHopf{{{\cal A}}} 
\def\augmentedh{{k}} 

\def\rescaledH{{H}} 
\def\rescaledAugH{{K}} 

\def\gIIA{g_{\text{\tiny IIA}}}
\def\tgIIA{\tilde{g}_{\text{\tiny IIA}}}
\def\gIIB{g_{\text{\tiny IIB}}}
\def\tgIIB{\tilde{g}_{\text{\tiny IIB}}}

\def\gYM{g_{\text{\tiny YM}}} 
\newcommand\gYMDim[1]{g_{\text{\tiny YM{{#1}}}}} 
\def\RCurvature{{\cal R}} 

\def\lst{{\ell_s}} 
\def\lPlanck{{\ell_p}} 

\def\dualphi{{\tilde{\phi}}} 
\def\BfieldNS{{B^{\text{\tiny (NSNS)}}}} 

\newcommand\Op[2]{{{\cal O}^{({#1})}_{{#2}}}} 
\newcommand\SUSY[1]{${\cal N}={#1}$}  

\def\Spin{{\text{Spin}}} 

\def\fThermal{{f}} 

\def\Area{{A}} 
\def\Volume{{\cal V}} 
\def\tVolume{{\tilde{\Volume}}} 

\def\VExpression{{V}} 

\def\dualizedRescaledU{{\tilde{\rescaledU}}} 
\def\dualizedRescaledH{{\tilde{\rescaledH}}} 
\def\dualizedRescaledAugH{{\tilde{\rescaledAugH}}} 
\def\dualizedDelta{{\tilde{\Delta}}} 

\def\tz{{\tilde{z}}} 

\def\angularForm{{\omega}}
\def\RCurrent{{\cal J}} 

\begin{document}
\begin{titlepage}
\begin{flushright}
hep-th/0702030\\
MAD-TH-07-01\\
UCB-PTH-07/02\\
LBNL-62285
\end{flushright}

\vfil\

\begin{center}

{\Large{\bf Aspects of Puff Field Theory}}

\vfil

Ori J. Ganor,$^{1,2}$ Akikazu Hashimoto,$^3$
Sharon Jue,$^1$\\
Bom Soo Kim,$^{1,2}$ and
Anthony Ndirango$^{1}$

\vfil

${}^1$Department of Physics, University of California, Berkeley, CA 94720\\
${}^2$Theoretical Physics Group, Lawrence Berkeley National Laboratory, Berkeley, CA 94720\\
${}^3$Department of Physics, University of Wisconsin, Madison, WI 53706
\vfil

\end{center}

\begin{abstract}
\noindent 
We describe some features of the recently constructed ``Puff Field
Theory,'' and present arguments in favor of it being a field theory
decoupled from gravity.  We construct its supergravity dual and
calculate the entropy of this theory in the limit of large 't Hooft
coupling.  We also determine the leading irrelevant operator that
governs its deviation from ${\cal N}=4$ super Yang-Mills theory.

\end{abstract}
\vspace{0.5in}

\end{titlepage}
\renewcommand{\baselinestretch}{1.05}  

\section{Introduction}
The Melvin Universe is an exact solution of 
Einstein gravity coupled with
gauge fields \cite{Melvin:1963qx}. It describes a consistent
gravitational backreaction when one attempts to support a uniform
magnetic field in the background. (Of course, the magnetic field is no
longer uniform when the gravitational backreaction is taken into
account.)
Melvin Universes are especially  natural  in the context of
Kaluza-Klein theory. Simply consider twisting an angular coordinate so that the space-time, where a plane is expressed in cylindrical coordinates, has the form
\be ds^2 = -dt^2 + dx^2 + dr^2 
+ r^2 (d \anglephi + \eta dz)^2 + dz^2 \ . \label{twist}\ee
Here $z \sim z + 2 \pi R$ is periodic, 
and $\eta$ therefore cannot be trivially eliminated by
a change of coordinates
(since a redefinition $\anglephi\rightarrow\anglephi+\eta z$
would modify the periodicity conditions on $z$ and $\anglephi$).
For $\eta = 0$, the space reduces to flat space in 4+1
dimensions. Kaluza-Klein reduction along the $z$ coordinate 
gives rise to a space-time with background magnetic field 
and some background scalar field configuration.

This type of space-time has a natural embedding in string theory. Simply embed (\ref{twist}) in 10 or 11 dimensional supergravity. One concrete realization is to embed (\ref{twist}) in type IIA supergravity. In this case, one can find an interesting type IIB supergravity solution by T-dualizing along the $z$ direction. The type IIB supergravity solution is of the form
\beq ds^2 & = & -dt^2 + d \vec x^2 + dr^2 
+ {{r^2 d \anglephi^2} \over {1 + \eta^2 r^2}}
+ {1 \over 1 + \eta^2 r^2} d\tilde z^2 \cr
B & = & 
  {\eta r^2 \over {1 + \eta^2 r^2}} 
     d \anglephi \wedge d \tilde z \cr
e^{\dilaton} & = & \sqrt{1 \over 1 + \eta^2 r^2} \cr
\tilde z &\sim& \tilde z + 2 \pi \tilde R, 
\qquad 
\tilde R  = {\alpha' \over R} \ .  
\label{melvinbg}
\eeq
This background is not supersymmetric. However, it is straightforward
to twist in more than one plane in such a way that some fraction of
supersymmetry is preserved.

These background space-times, from the point of view of string theory,
are special in that their world sheet sigma model is exactly solvable
\cite{
Russo:1994cv,Tseytlin:1994ei,Russo:1995tj,Tseytlin:1995fh,Russo:1995aj,Russo:1995ik}. This
follows, in essence, from the fact that they are dual to flat space
with some periodic identifications.

Melvin universes are also useful for constructing non-local quantum
field theories as a decoupling limit. Simply consider adding a
D3-brane in the background (\ref{melvinbg}) and take the appropriate
scaling limit for the parameters $\eta$, $r$, and $R$ as one sends
$\alpha'$ to zero. In enumerating distinct embeddings of D3 branes in this background, 
it is useful to note that there are essentially two
special spatial coordinates, $\tilde z$ and $\anglephi$. 
Taking both
$\tilde z$ and $\anglephi$ to be along the brane gives rise to
non-commutative geometry with a position dependent non-commutativity
parameter \cite{Hashimoto:2004pb,Hashimoto:2005hy} for which the
deformation quantization formula of Kontsevich
\cite{Kontsevich:1997vb} becomes relevant.
Taking $\tilde z$ to be along the brane but $\anglephi$ 
to be transverse gives rise to a dipole
deformation, which was introduced in
\cite{Bergman:2000cw,Bergman:2001rw}.  
The construction by Lunin and
Maldacena of $\beta$-deformed \SUSY{4} superconformal theory 
\cite{Lunin:2005jy}
can also be viewed as an example of this construction where both 
$\tilde z$ and $\anglephi$ are transverse to the brane. 
Other constructions of a similar kind were studied in 
\cite{Hashimoto:2002nr} and \cite{Alishahiha:2003ru}.
There are
generalizations of these constructions that can be obtained
by modifying the embeddings of
the $\anglephi$ and the $\tilde z$ coordinates in the full geometry, which
were classified and tabulated in \cite{Hashimoto:2004pb}. 
Most of these
constructions give rise to a non-local field theory as a decoupling
limit of string theories with solvable world sheet sigma model.

Recently, one of us proposed an example of a non-local field theory
not included in \cite{Hashimoto:2004pb}, which arises from a novel
embedding of a D-brane in a Melvin universe \cite{Ganor:2006ub}. 
The construction proceeds as follows. 
\begin{enumerate}
\item Start with $N$ D0-branes in flat 10
dimensional background of type IIA string theory. 
The M-theory lift of
this configuration corresponds to a Kaluza-Klein wave 
traveling along the M-theory circle. 
Let us parameterize the coordinates of the background
geometry in 11 dimensions using coordinates
\be ds^2 = -dt^2 + \sum_{i=1}^3 dx_i^2
+ \sum_{i=1}^3 (dr_i^2 + r_i^2 d \anglephi_i^2) + dz^2,
\qquad 
z \sim z + 2 \pi R \ . \ee
\item Twist the angular coordinate $\anglephi_i$ by a deformation parameter $\beta_i$ so that the metric becomes
\be ds^2 = -dt^2 + \sum_{i=1}^3 dx_i^2
+ \sum_{i=1}^3 \left\lbrack
dr_i^2 + r_i^2 (d \anglephi_i + \beta_i dz)^2\right\rbrack
+ dz^2\ . 
\ee
\item Reduce back to type IIA along the $z$ coordinate, giving rise to a D0-brane in a Melvin universe 
supported by a flux of magnetic RR 2-form field strength.

\item T-dualize along the $x_1$, $x_2$, and $x_3$ directions. This will give rise to  a configuration of D3-branes embedded in a Melvin universe with  RR 5-form field strength in the background in type IIB theory.
\end{enumerate}
The construction described in \cite{Ganor:2006ub} started with a KK
wave in type II theory but leads to the same geometry. 

By taking a suitable scaling limit involving $\alpha'$, $\beta_i$,
and
the compactification radii as we will describe in more detail
below, one arrives at a presumably decoupled system of 
non-local quantum field theory, 
similar in many regards to non-commutative Yang-Mills theory,
dipole theory, and NCOS. This theory was named the ``Puff Field
Theory'' (PFT) in \cite{Ganor:2006ub}, because the light degrees of
freedom ``puff up'' in all three dimensions. The distinguishing
feature of PFT is the fact that it leaves unbroken a spatial
subgroup of the Lorentz group $SO(3) \subset SO(1,3)$, 
unlike the more familiar
non-commutative/dipole field theories. Such a theory could possibly be
of phenomenological interest when applied to maximally isometric
cosmological scenarios of the Freedman-Robertson-Walker type.

The goal of this article is to describe the features of PFT
and to provide more evidence in support of the conjecture
that PFT is decoupled from gravity.
This is facilitated by the explicit construction 
of the supergravity dual.
This paper is organized as follows. 
In section~\ref{sec:SupergravityDual}, we
describe the construction of the supergravity dual itself. 
In section~\ref{sec:Thermodynamics}, 
we construct the finite temperature case and describe the
thermodynamics of PFT.
In section~\ref{sec:Renormalization},
we describe the RG flow of PFT in greater detail.
In section~\ref{sec:Deformation},
we identify the leading irrelevant
operator responsible for deforming \SUSY{4} SYM to PFT.  
We conclude in section~\ref{sec:Concluding}.

\section{Supergravity dual of PFT
\label{sec:SupergravityDual}}

In this section, we describe the construction of the supergravity solution that is dual to PFT. 
In what follows, we present the steps leading 
to the solution (\ref{eqn:dualSG}) in some detail.
After that, we will analyze the regime of validity of
the classical supergravity solution.

To obtain the supergravity dual of PFT,
we start from type IIA supergravity compactified along the $x_i$ directions and consider $N$ D0-branes smeared along the compact directions. In anticipation of the $T$-duality, we will denote the period of compactification of $x_i$ by 
${\alpha' /  R_i}$. 
This geometry can be written explicitly in the form
\beq ds^2 &=& -h^{-1/2} dt^2 + h^{1/2} 
\left(\sum_{i=1}^3 dx_i^2 + \sum_{i=4}^9 d y_i^2 \right) \cr
A & = & h^{-1} dt \cr
e^{2 \dilaton} &=& h^{3/2}\ .  \label{IIA} \eeq
where
\be h = 1 + 
{60 \pi^3 \gIIA N \alpha'^{7/2} \over (\|x\|^2 + \|y\|^2)^{7/2}}
\ee
for the ``non-smeared'' solution (for which directions
$1,2,3$ are noncompact and $-\infty<x_1, x_2, x_3<\infty$),
and
\be h = 1 + {4 \pi \gIIA N R_1 R_2 R_3   \alpha'^{1 / 2} \over 
\|y\|^4} \ee
for the ``smeared'' solution (which is obtained
from the ``non-smeared'' one
by replacing the second term in $h$ with its integral
over the full range of $x_1, x_2, x_3$ and dividing by the
total volume $(2\pi\alpha')^3/R_1 R_2 R_3$ of the $T^3$).
Here 
\be
\|x\|^2\equiv\sum_{i=1}^3 x_i^2,
\qquad
\|y\|^2\equiv\sum_{i=4}^9 y_i^2\ .
\ee
Now we perform the M-theory lift, twist, and reduction back to type
IIA.  The M-theory lift of (\ref{IIA}) is purely geometric
\be ds^2 =  - h^{-1} dt^2 + h (dz - h^{-1} dt)^2 +  
\left(\sum_{i=1}^3 dx_i^2 + \sum_{i=4}^9 d y_i^2 \right)\ . \ee
Here $z$ is a periodic coordinate (the ``M-circle'')
with period $z\sim z+2\pi \gIIA\lst$ 
(where $\lst\equiv{\alpha'}^{1/2}$ is the string scale).
One can in principle consider letting all $\beta_i$'s take
independent values in performing the twist.  We will, however,
concentrate on the case where $\beta_1 = \beta_2 = \eta$ and
$\beta_3 =0$, which leaves half of the supersymmetries unbroken,
and will at the end give us a theory with ${\cal N}=2$ in 4D.
Then, the metric after the twist has the form
\be ds^2 =  - h^{-1} dt^2 + h (dz - h^{-1} dt)^2 +  
\sum_{i=1}^3 dx_i^2 + \sum_{i=1}^2
\left\lbrack dr_i^2  
+ r_i^2 (d \anglephi_i + \eta dz)^2\right\rbrack
 + \sum_{i=8}^9 d y_i^2\ . 
\label{eqn:twisted}
\ee
Before we proceed, we make another convenient change of variables.
We replace the four coordinates $r_1, r_2, \anglephi_1, \anglephi_2$
with the radial variable
\be
\fullu\equiv \sqrt{r_1^2 + r_2^2},
\qquad
(0\le\fullu<\infty)\ ,
\ee
and three new angular coordinates
$\fiberphi, \basetheta, \basephi$ 
defined as follows,
\be
\basephi\equiv\anglephi_1-\anglephi_2,
\qquad
\sin\basetheta\equiv \frac{2 r_1 r_2}{r_1^2 + r_2^2},
\qquad
\fiberphi\equiv\anglephi_1\ .
\ee
For fixed $t,x_1,x_2,x_3,y_8,y_8,$ and $\fullu$,
the variables $\fiberphi, \basetheta, \basephi$
describe an $S^3$ in the form of a Hopf fibration:
$(\basetheta,\basephi)$  are spherical coordinates on the
$S^2$ base, and $\fiberphi$ is a periodic coordinate
on the $S^1$ fiber with period $2\pi.$
Presenting $S^3$ as a Hopf fibration is convenient
because the twist is in the direction of the fiber $\fiberphi.$
In order to save space,
we denote the (Fubini-Study)
metric on the base of the Hopf fibration by
\be
\dsBase{2}^2\equiv\frac{1}{4}
(d\basetheta^2 + \sin^2\basetheta\,d\basephi^2)\ ,
\label{eqn:dsBase}
\ee
and we denote the connection of the Hopf fibration by
\be
\connHopf\equiv
-\frac{1}{2}(1-\cos\basetheta)d\basephi\ .
\label{eqn:defconnHopf}
\ee
The metric can now be expressed in the form
\beq ds^2 &=&  
- h^{-1} dt^2 + h (dz - h^{-1} dt)^2 
+ \sum_{i=1}^3 dx_i^2 
+ d\fullu^2
\cr &&\qquad
+ \fullu^2 \left\lbrack
\dsBase{2}^2 
+ (d \fiberphi +  \eta dz + \connHopf)^2\right\rbrack 
+ \sum_{i=8}^9 d y_i^2\ .
\label{Mtheory} 
\eeq
Reducing along $dz$ to IIA then gives
\beq ds^2 & = & 
\augmentedh^{1/2}\Bigl( - h^{-1} dt^2
+\sum_i dx_i^2 
+ d\fullu^2  + \fullu^2 \dsBase{2}^2 
+\sum_{i=8,9} dy_i^2 \Bigr)
\cr && \qquad
+ \augmentedh^{-1/2}h\fullu^2
(d \fiberphi + \connHopf + \eta h^{-1} dt)^2\ ,
\cr
A & = &  
\augmentedh^{-1}
  \left( -  dt 
 + \fullu^2 \eta (d \fiberphi  + \connHopf) \right)\ , \cr
e^\dilaton & = & \gIIA \augmentedh^{3/4}\ ,
\eeq
where
\be
\augmentedh\equiv h + \eta^2 u^2 
=1 + {4 \pi \gIIA N R_1 R_2 R_3 \alpha'^{1/2} 
   \over (u^2 + \|y\|^2)^2} + \eta^2 u^2\ .
\ee
T-dualizing along $x_1$, $x_2$, and $x_3$ then gives rise to the supergravity solution
\beq ds^2 & = & 
\augmentedh^{1/2} \left( - h^{-1} dt^2
+ d\fullu^2  + \fullu^2 \dsBase{2}^2 
+\sum_{i=8,9} dy_i^2 \right) 
\cr && \qquad
+\augmentedh^{-1/2} \left(
\sum_i dx_i^2 
+ h  \fullu^2(d \fiberphi + \connHopf + \eta h^{-1} dt)^2
\right)\ ,
\cr
A & = &  \augmentedh^{-1}
\left( -  dt + \fullu^2 \eta (d \fiberphi+\connHopf) \right) 
\wedge dx_1 \wedge dx_2 \wedge dx_3\ ,
\cr
e^\dilaton & = & \gIIB =  R_1 R_2  R_3 \alpha'^{-3/2} \gIIA\ .
\label{eqn:FullSolution}
\eeq
Here $A$ is not quite the full Ramond-Ramond gauge 4-form field,
but the full Ramond-Ramond 5-form field strength is
the self-dual part of $dA.$
In the new variables, $h$ and $\augmentedh$ take the forms
\be h = 1 + {4 \pi \gIIB N \alpha'^2 \over 
(\fullu^2 + \|y\|^2)^2} \ ,\qquad
\augmentedh = 1 + {4 \pi \gIIB N \alpha'^2 \over 
(\fullu^2 + \|y\|^2)^2} +\eta^2 \fullu^2\ .
\label{eqn:hh} \ee
Note that if we set $N=0$ we get $h=1$
and the supergravity solution reduces to
a Melvin universe with background RR 5-form flux.
For $N\neq 0$, 
The warping due to $h(\fullu,y)$ describes the gravitational 
back-reaction of the D3-branes.

Now, we can take the decoupling limit following the procedure of \cite{Maldacena:1997re}. 
As usual, in order to decouple the gauge theory from
the rest of the string theory modes, we take the zero slope
limit $\alpha'\rightarrow 0$,
and we need to specify how to scale the coordinates
$y_8, y_9,\fullu$ and the twist parameter $\eta$ in this
limit.
It turns out that the appropriate scaling is to keep finite
the following rescaled coordinates 
\be
\rescaledU\equiv {\alpha'}^{-1}\fullu,
\qquad
\rescaledY_i\equiv {\alpha'}^{-1}y_i\quad (i=8,9)
\ ,
\ee
while scaling the twist parameter $\eta$ so that
\be \Delta^3 \equiv \eta \alpha'^2 = \text{fixed}\ , 
\label{eqn:defDelta}
\ee
and keeping $\gIIB = 2 \pi \gYM^2$, $R_1$, $R_2$,
and $R_3$ fixed. 
This scaling is chosen so that in the dual supergravity
solution the effects of the deformation will be finite
in the scaling limit. For example, the second and
third terms in $\augmentedh$ in (\ref{eqn:hh}) are
comparable in this scaling limit.
Our scaling limit also
turns out to be the one suggested in \cite{Ganor:2006ub}
using different arguments.

 Note in particular that this is the
scaling that keeps the angle of twist 
per radius of the M-circle,
\be \chi \equiv \gIIA\lst\eta = 
{\eta \gIIB \alpha'^{2} \over  R_1  R_2  R_3} 
= {2\pi \gYM^2 \Delta^3 \over  R_1  R_2  R_3}\ ,\ee
finite. We will see below that an integer shift in $\chi$ leads to an equivalent theory up to a certain duality, similarly to the structure of Morita equivalence encountered in the context of ordinary non-commutative field theories.
We are now almost ready to take the $\alpha'\rightarrow 0$ limit.
We define the scaled harmonic functions 
(with the notation 
\textit{$\|\rescaledY\|^2\equiv \rescaledY_8^2 + \rescaledY_9^2$})
\be 
\rescaledH\equiv\lim_{\alpha'\rightarrow 0}\alpha'^2 h 
={4 \pi \gIIB N \over (\rescaledU^2 + \|\rescaledY\|^2)^2}\ ,
\qquad
\rescaledAugH\equiv\lim_{\alpha'\rightarrow 0}\alpha'^2\augmentedh
={4 \pi \gIIB N \over (\rescaledU^2 + \|\rescaledY\|^2)^2} 
+ \Delta^6 \rescaledU^2\ ,
\label{eqn:defrescaledHAugH}
\ee
which captures the asymptotic behavior of the harmonic functions
$h$ and $\augmentedh$ in the decoupling limit, 
and depend only on the PFT parameters
\textit{$\gIIB\equiv 2\pi\gYM^2$} and $\Delta.$
In terms of $\rescaledH$ and $\rescaledAugH$ we can write
\beq {ds^2\over \alpha'} & = & 
\rescaledAugH^{1/2} \left( - \rescaledH^{-1} dt^2
+ d\rescaledU^2  + \rescaledU^2 \dsBase{2}^2 
+\sum_{i=8,9} d\rescaledY_i^2 \right) 
\cr && \qquad
+\rescaledAugH^{-1/2} \left(
\sum_i dx_i^2 
+ \rescaledH\rescaledU^2
(d \fiberphi + \connHopf + \Delta^3\rescaledH^{-1} dt)^2
\right)\ ,
\cr
{A \over \alpha'^2} & = &  
\rescaledAugH^{-1}
\left( -  dt + \rescaledU^2 \Delta^3 
(d \fiberphi + {\cal A}) \right) 
\wedge dx_1 \wedge dx_2 \wedge dx_3\ ,
\cr
e^\dilaton & = & \gIIB = 2 \pi \gYM^2\ . \label{eqn:dualSG}
\eeq
This is an exact solution of the classical
equations of motion of type IIB supergravity,
and we will interpret it as the supergravity dual of PFT.  
Much of the conclusions we draw regarding the nature of PFT 
will be based on this supergravity solution, 
which is one of the main results we are
reporting in this paper.

PFT depends on a dimensionful parameter $\Delta$,
which according to its definition in (\ref{eqn:defDelta}),
has dimensions of length.
For generic $\Delta$, the solution (\ref{eqn:dualSG})
is invariant under Poincar\'e transformations in
the $t, x_1, x_2, x_3$ directions, under $SU(2)$
rotations of the base $B(2)$ of the Hopf fibration
(acting on the spin structure and the fiber direction
$\fiberphi$ as well), under $U(1)$ translations generated by
the vector field $\partial/\partial\fiberphi$, 
and under 8 supersymmetries.
For $\Delta=0$ the solution (\ref{eqn:dualSG})
reduces to $AdS_5\times S^5$ --
the coordinates $t, x_1, x_2, x_3$ and 
\be
\radialV\equiv\sqrt{\rescaledU^2 +\|\rescaledY\|^2}
\label{eqn:defV}
\ee
parameterize the $AdS_5$ part.
($\radialV$ can be traced back to the radial direction
transverse to the D3-brane.)
For later use, it is also convenient to define
the angle $0\le\anglepsi\le\pi/2$ by
\be
\rescaledU=\radialV\cos\anglepsi,\qquad
\|\rescaledY\|=\radialV\sin\anglepsi\ .
\label{eqn:defchi}
\ee
Up to factors of certain powers of 
$\lambda = 2 \pi \gYM^2 N = \gIIB N$
that will be discussed later,
$\Delta^{-1}$ sets the interesting energy scale for PFT.
This is the scale above which
PFT becomes appreciably different from \SUSY{4} SYM. 
This can be seen directly from (\ref{eqn:dualSG}):
for $\rescaledU \ll \Delta^{-1}$ (and fixed $\lambda$), the
solution asymptotes to $AdS_5 \times S_5$, indicating that the
infra-red fixed point of this theory is \SUSY{4} SYM.
On the other hand, for $\rescaledU \gg \Delta^{-1}$, 
the solution deviates strongly from the
$AdS_5\times S_5$ background. The supergravity duals of
other nonlocal field theories such as 
non-commutative Yang-Mills theory
\cite{Hashimoto:1999ut,Maldacena:1999mh} and dipole theory
\cite{Bergman:2001rw} also exhibit similar features.

\subsubsection*{Regime of validity}

Let us comment on the region of validity of the dual supergravity description. 
The 't Hooft coupling constant
$\lambda = 2 \pi \gYM^2 N = \gIIB N$ must be large in order for the curvature to be weak.
We assume that the Yang-Mills coupling constant $\gYM$ itself
is kept finite.
(Of course, $\gYM$ has to be small if one wishes to
extend the discussion beyond the classical supergravity
description, for example, to include the excited string spectrum.)

If $\lambda\gg 1$,
the curvature is small everywhere
provided $\anglepsi\neq \pi/2.$
More specifically,
the invariant square of the curvature tensor,
as calculated from the string-frame metric (\ref{eqn:dualSG}), is
\beq
\RCurvature_{\mu\nu\sigma\tau}\RCurvature^{\mu\nu\sigma\tau} &=&
{\alpha'}^{-2}
\bigl(
4\pi\lambda
+\Delta^6 \rescaledU^2\radialV^4
\bigr)^{-5}
\Bigl\lbrack
80(4\pi\lambda)^4
-80(4\pi\lambda)^3\Delta^6\radialV^4 (3\rescaledU^2 + 5\radialV^2)
\cr &&
+24(4\pi\lambda)^2\Delta^{12}\radialV^8
  (136\rescaledU^4 + 29\rescaledU^2\radialV^2 + 5\radialV^4)
\cr &&
+32\pi\lambda\Delta^{18}\radialV^{12}\rescaledU^2
  (15\radialV^4 + 7\rescaledU^2\radialV^2 -72\rescaledU^4)
+65\Delta^{24} \rescaledU^4\radialV^{20}
\Bigr\rbrack\ .
\eeq
So, if $\anglepsi\neq\pi/2$ we see that for
\textit{$\rescaledU\Delta\ll\lambda^{1/6}$} the curvature
scale is of order \textit{${\alpha'}^{-1/2}\lambda^{-1/4}$},
while for \textit{$\rescaledU\Delta\gg\lambda^{1/6}$}
the curvature scale is of order
\textit{${\alpha'}^{-1/2}(\rescaledU\Delta)^{-3/2}$}. 
Both of these
quantities are small for $\lambda \gg 1$.
If $\anglepsi=\pi/2$, on the other hand, the curvature is small
only for  \textit{$\Delta\|\rescaledY\|\ll\lambda^{3/4}$}. Therefore, 
observables that sensitively probe the $\anglepsi=\pi/2$ region might receive corrections due to stringy effects.

Another requirement for the classical supergravity analysis
to be applicable is that the proper size of the various
compact directions be large compared to the string scale.
For the $\fiberphi$ direction we get
\be
\rescaledAugH^{-1/2}\rescaledH\rescaledU^2 \gg 1.
\ee
Assuming again that $\anglepsi\neq\pi/2$,
we find that the $\fiberphi$ direction is large if
\textit{$\rescaledU\Delta\ll\lambda^{1/3}$}.
On the other hand, if \textit{$\rescaledU\Delta\gg\lambda^{1/3}$}
the $\fiberphi$-circle is smaller than string scale
and its radius is of order
\textit{${\alpha'}^{1/2}\lambda^{1/2}(\rescaledU\Delta)^{-3/2}$}.
In this regime 
the supergravity dual (\ref{eqn:dualSG}) cannot be trusted.
For $\anglepsi\neq\pi/2$, the radius of the $\fiberphi$-circle
shrinks to zero, but the solution is not singular.
To see this, note that for fixed nonzero $\|\rescaledY\|$ we
have $\radialV\neq 0$, and as 
$\anglepsi\neq\pi/2$ (and therefore $\rescaledU\rightarrow 0$)
the metric on the base \textit{$\dsBase{2}^2$} and
the fiber $\fiberphi$ combine to a metric on $S^3$, and together
with the $\rescaledU$ direction we get the metric on a ball.

Next, we need to discuss directions $x_1, x_2, x_3.$
The proper size of each of these compact directions
needs to be large in comparison to string scale. 
One way to achieve this is to simply take the 
decompactification limit $R_i\rightarrow\infty$ ($i=1,2,3$)
and formulate PFT on $\Reals^{3,1}$, so to speak.
Alternatively, we can keep the compactification radii
$R_1, R_2, R_3$ finite.
Taking this approach, as we shall see in 
section~\ref{sec:Renormalization}, yields a richer
structure of energy scales in the theory,
but then in order 
for the classical supergravity solution (\ref{eqn:dualSG})
to be valid, we need the additional requirements
\be
R_i\gg\rescaledAugH^{1/4},\qquad i=1,2,3.
\label{eqn:RiA}
\ee
Assuming that $\anglepsi\neq\pi/2$, 
we find that \textit{$\rescaledAugH$}
is of the order of 
\textit{$4\pi\lambda\rescaledU^{-4}+\Delta^6\rescaledU^2$}.
This expression is never smaller than
\textit{$3(\pi\lambda)^{1/3}\Delta^4$}, and therefore 
(\ref{eqn:RiA}) will not be satisfied unless
\be  R_i \gg \lambda^{1/12} \Delta\ ,
\qquad (i=1,2,3). \label{eqn:assump} \ee
Assuming (\ref{eqn:assump}) now,
the condition (\ref{eqn:RiA}) sets the following range
requirement for $\rescaledU$:
\be {\lambda^{1/4} \over  R_i} \ll 
\rescaledU \ll { R_i^2  \over \Delta^3} \ . \label{eqn:Urange} \ee
The validity conditions that we found so far can be recast in
terms of the energy scale.
For $AdS_5\times S^5$,
the holographic energy/distance relation  \cite{Peet:1998wn}
takes the form
\be E = {\rescaledU \over \sqrt{\lambda}} \ . \label{uvir} \ee
In order of magnitude,
this form is also applicable to our metric, at least 
if we assume that $\cos\anglepsi$ is of order $O(1)$,
so that $\rescaledU$ and $\radialV$ are comparable.
We will demonstrate this later in (\ref{eqn:UT}) of
section~\ref{sec:Thermodynamics}, 
when we discuss PFT at nonzero temperature.

The conditions  (\ref{eqn:Urange}) can now be written
as a range of energy scales
\be {\lambda^{-1/4} \over  R_i} \ll E \ll {\lambda^{-1/2}  R_i^2  \over \Delta^3} \ , \label{eqn:Erange} \ee
and the condition about the $\fiberphi$-circle
that was found above becomes
\be
E\ll\frac{\lambda^{-1/6}}{\Delta}
\label{eqn:ERangeFiber}.
\ee
Combining (\ref{eqn:ERangeFiber}) and (\ref{eqn:Erange}) we get
\be 
{\lambda^{-1/4} \over  R_i} \ll E \ll 
\min\bigl\{
\frac{\lambda^{-1/6}}{\Delta},
{\lambda^{-1/2}  R_i^2  \over \Delta^3}
\bigr\} \ . \label{eqn:ERangeAll} 
\ee
Similar sets of bounds on the region of validity for the case of non-commutative Yang-Mills theory were 
pointed out in \cite{Hashimoto:1999yj}.
Note also that, assuming (\ref{eqn:assump}) and $\lambda\gg 1$,
we have the inequality
\be 
{\lambda^{-1/4} \over  R_i}
\ll
\frac{\lambda^{-1/3}}{\Delta}
\ll 
\min\bigl\{
\frac{\lambda^{-1/6}}{\Delta},
{\lambda^{-1/2}  R_i^2  \over \Delta^3}
\bigr\} \ . 
\label{eqn:InterestingScale}
\ee
The energy scale \textit{$\lambda^{-1/3}/\Delta$} is 
important because it corresponds to 
\textit{$\rescaledU\Delta = \lambda^{1/6}$},
which is the scale at which the metric starts to deviate
markedly from $AdS_5\times S^5.$ 
For example, below that scale $\rescaledH\approx\rescaledAugH.$
This is therefore the scale at which PFT effects enter into
play, and we see from (\ref{eqn:InterestingScale})
that it is inside the range of validity (\ref{eqn:ERangeAll}).

In (\ref{eqn:ERangeAll}),
the lower bound \textit{$\lambda^{-1/4}/R_i \ll E$} is
independent of the non-locality and applies 
just as well to the case of ordinary $AdS_5 \times S_5$
compactified on a circle. The bound simply
indicates the presence of finite size effects cutting off the spectrum in the IR. 
The order of magnitude of the
size of a typical excitation with energy $E$ can be
estimated as the Compton length $L = 1/E$, 
but this estimate fails
when the size of the excitation gets bigger 
than the size of the box.
This explains why the lower bound on $E$ is proportional to
$1/R_i.$
The factor of \textit{$\lambda^{-1/4}$} in the bound
is the effect of strong coupling.
For energies below the bound,
\textit{$E\ll\lambda^{-1/4}/R_i$}, one should
look for a description in terms of the near horizon 
geometry of a lower dimensional brane. Readers are referred to
\cite{Itzhaki:1998dd,Brandhuber:1998er} for explanations concerning
the correct cross-over behavior and the correspondence 
principle at work around this scale.

The upper bound on $E$ in
(\ref{eqn:Erange}) implies that the size of a typical excitation
starts to grow with energy according to 
\textit{$L\sim\sqrt{\lambda^{1/2}\Delta^3 E}$},
so that the upper bound is reached when \textit{$L\sim R_i$}.
This is a characteristic feature of non-local field theories. 
The size of an object grows both 
in the extreme IR and in the extreme UV.
When the size of the
object becomes larger than the size of the box, 
one must adopt an alternative description. 
We will comment further on this issue in
section~\ref{sec:Renormalization}.

\subsubsection*{The high energy regime 
\textit{$E\gg \lambda^{-1/6}/\Delta$}}

So far, we discussed the upper bound on energy coming from
(\ref{eqn:Erange}), but we also have another upper bound
from (\ref{eqn:ERangeFiber}).
The latter suggests another interesting 
length-scale in the problem,
namely \textit{$\lambda^{1/6}\Delta$},
at least for strong `t Hooft coupling $\lambda\gg 1.$
As we reach the corresponding range 
\textit{$\rescaledU\sim\lambda^{1/3}/\Delta$}
in the supergravity solution (\ref{eqn:dualSG})
the ten-dimensional description loses its classical
interpretation.
If the compactification radii $R_1,$ $R_2$ and $R_3$
are all much bigger than the length scale
\textit{$\lambda^{1/6}\Delta$}, then
\be
\frac{\lambda^{1/3}}{\Delta}
\ll
\frac{R_i^2}{\Delta^3},
\ee
and the energy scale $E$ (or equivalently, $\rescaledU$)
at which the $\fiberphi$-circle becomes comparable to
the string length-scale is lower than the energy scale
at which the radii of the $x_1$, $x_2$, $x_3$ fall below
the string scale.
In particular, this is the case in the decompactification limit
$R_i\rightarrow\infty$.
There is then a range of $\rescaledU$ for which,
even though the $\fiberphi$-circle is small,
we can still dimensionally reduce
along it to get a valid nine-dimensional
classical supergravity description, as long as we keep
away from the $\anglepsi=\pi/2$ locus, near which
the size of the fiber varies rapidly.

In the extreme regime 
\textit{$\rescaledU\gg\lambda^{1/3}/\Delta$}
the $\fiberphi$-circle is smaller than string scale,
and it therefore makes sense to apply a T-duality transformation,
at least away from the $\anglepsi=\pi/2$ locus.
Using the formulas of \cite{Bergshoeff:1995as} we arrive
at a background with the following NSNS fields:
\beq {ds^2\over \alpha'} & = & 
\rescaledAugH^{1/2} \left( - \rescaledH^{-1} dt^2
+ d\rescaledU^2  + \rescaledU^2 \dsBase{2}^2 
+\sum_{i=8,9} d\rescaledY_i^2 \right) 
\cr && \qquad
+\rescaledAugH^{-1/2}\sum_i dx_i^2 
+\rescaledAugH^{1/2}\rescaledH^{-1}\rescaledU^{-2}d \dualphi^2\ ,
\cr
\frac{1}{\alpha'}\BfieldNS &=& d\dualphi\wedge
(\connHopf + \Delta^3\rescaledH^{-1} dt)\ ,
\cr
e^\dilaton & = & 
2 \pi \gYM^2 \rescaledAugH^{1/4}\rescaledH^{-1/2}\rescaledU^{-1}\ . 
\label{eqn:dualSGIIA}
\eeq
Here $\dualphi$ is a periodic variable with period $2\pi$
that parameterizes the T-dual circle, and
there are also nonzero RR fields that 
have not been written down here, for simplicity.
(See also \cite{Duff:1998us} for a related discussion
where T-duality has been applied to $AdS_5\times S^5$
by viewing the $S^5$ as a Hopf fibration over a base 
\textit{$CP^2$}.)
There are, however, at least three extra complications:
\begin{enumerate}
\item
The physics at the 
vicinity of the locus $\anglepsi=\pi/2$ is not 
captured properly by (\ref{eqn:dualSGIIA}).
As we will now explain, the strongly curved metric in that
region should be replaced by an NS5-brane.
To see this, first note that for fixed $t$,
$x_1$, $x_2$, $x_3$, $\dualphi$ and fixed $\radialV\neq 0$,
the remaining parts of the metric
\be
\rescaledAugH^{1/2} \left( 
d\rescaledU^2  + \rescaledU^2 \dsBase{2}^2 
+\sum_{i=8,9} d\rescaledY_i^2 \right) 
\label{eqn:pieceds}
\ee
describe a space that is topologically equivalent to
an \textit{$S^4$}.
In fact, defining a new periodic coordinate
$0\le\angleY<2\pi$ by
\be
\rescaledY_8 = \|\rescaledY\|\cos\angleY,
\qquad
\rescaledY_8 = \|\rescaledY\|\sin\angleY,
\ee
our piece of the metric reduces to
\be
\rescaledAugH^{1/2}
\radialV^2 \bigl(
d\anglepsi^2 
+\frac{1}{4}\cos^2\anglepsi\,
(d\basetheta^2 + \sin^2\basetheta\,d\basephi^2)
+\sin^2\anglepsi\,d\angleY^2\bigr)\ ,
\label{eqn:almostS4}
\ee
where we used (\ref{eqn:dsBase}).
And were it not for the explicit dependence of 
$\rescaledAugH$ on $\anglepsi$ and 
for the factor of $1/4$ in front of the second term,
the metric (\ref{eqn:almostS4}) would describe an $S^4$ exactly.
We are now ready to analyze the region near $\anglepsi=\pi/2.$
The locus $\anglepsi=\pi/2$ is an $S^1$ 
(that can be parameterized by $\angleY$), and
the $(\basetheta,\basephi)$ variables describe an \textit{$S^2$}
that shrinks to zero as $\anglepsi\rightarrow \pi/2.$
The radius of the $\dualphi$-circle, on the other hand,
increases indefinitely.
As we approach $\anglepsi=\pi/2$, however, the flux of the NSNS 
3-form field-strength through the \textit{$S^2\times S^1$}
(generated by $\basetheta$, $\basephi$ and $\dualphi$)
remains finite:
$$
\frac{1}{\alpha'}\int_{S^2\times S^1}d\BfieldNS =
 d\dualphi\wedge d\connHopf =  4\pi^2.
$$
This indicates the presence of one unit of NS5-brane charge
at $\anglepsi=\pi/2$.

\item
When applying T-duality in superstring theory to a background
that is a circle fibration, one has to be careful about
the boundary conditions for fermions along the fiber direction.
Specifically, before the T-duality, consider
the holonomy for a closed path that wraps the fiber over
a fixed point in the base. Assuming that the fiber varies
slowly over the base, the geometrical holonomy is close to
the identity in $SO(10)$, but when lifted to spinors
the holonomy could be close to either $(+1)$ or $(-1)$
in $\Spin(10).$
We can determine which case corresponds to our metric
by noting that the fiber of the Hopf fibration $S^1$
(the $\fiberphi$-circle) over $S^2$ 
(the $\basetheta,\basephi$ base) is contractible,
and its spin holonomy can therefore be calculated unambiguously,
and it is easy to see that it is $(-1).$
Note, however, that the full holonomy of fermions along the fiber
is $(+1),$ because our solution preserves supersymmetry.
The minus sign from the geometrical spin holonomy
is canceled by the nongeometrical contribution to the 
holonomy due to the RR 5-form field strength.

\item
Finally, we note that the dilaton in (\ref{eqn:dualSGIIA})
gets large for
\be
\rescaledU\sim\frac{\lambda^{1/3}}{\gIIB^{2/3}\Delta}
 \sim\frac{N^{2/3}}{\lambda^{1/3}\Delta}\ ,
\label{eqn:StrongIIA}
\ee
assuming $\anglepsi\neq\pi/2$, as usual.
This scale of $\rescaledU$ is larger than the bound
\textit{$\lambda^{1/3}/\Delta$},
because we are always taking $N$ to be very large.
Thus, (\ref{eqn:dualSGIIA}) is likely to be valid in a range
\be
\lambda^{1/3}/\Delta\ll
\rescaledU\ll\frac{N^{2/3}}{\lambda^{1/3}\Delta}\ .
\ee
The upper bound (\ref{eqn:StrongIIA}) is smaller than
the upper bound of (\ref{eqn:Urange}) if
\be
\Delta < N^{-1/3}\lambda^{1/6}R_i\ .
\label{eqn:ConditionDelta}
\ee
For finite $R_i$, this is never the case, but
in the decompactification limit $R_i\rightarrow\infty$ this holds.
Then, for \textit{$\rescaledU\gg N^{2/3}\lambda^{-1/3}/\Delta$}
the dilaton becomes large, and a proper description
requires 11-dimensional supergravity.
Lifting the solution (\ref{eqn:dualSGIIA}) to M-theory,
we get the metric
\beq
\frac{ds^2}{\lPlanck^2} &=&
N^{2/3}\lambda^{-2/3}\Bigl\lbrack
\rescaledAugH^{1/3}\rescaledH^{1/3}\rescaledU^{2/3}
\bigl( - \rescaledH^{-1} dt^2
+ d\rescaledU^2  + \rescaledU^2 \dsBase{2}^2 
+\sum_{i=8,9} d\rescaledY_i^2 \bigr) 
\cr && \qquad
+\rescaledAugH^{-2/3}\rescaledH^{1/3}\rescaledU^{2/3}
\sum_i dx_i^2 
+\rescaledAugH^{1/3}\rescaledH^{-2/3}\rescaledU^{-4/3}
d \dualphi^2\Bigr\rbrack
\cr && \qquad
+N^{-4/3}\lambda^{4/3}
\rescaledAugH^{1/3}\rescaledH^{-2/3}\rescaledU^{-4/3} 
d\Mcircle^2
\ ,
\eeq
where $0\le \Mcircle<2\pi$ is a new periodic coordinate.
For fixed $\anglepsi\neq\pi/2$ and 
$\rescaledU\gg\lambda^{1/6}/\Delta$, we may approximate
\be
\rescaledH\approx 4\pi\lambda\rescaledU^{-4}\cos^4\anglepsi,
\qquad
\rescaledAugH\approx\Delta^6\rescaledU^2\ ,
\ee
and
\beq
\frac{ds^2}{\lPlanck^2} &\approx&
\left(\frac{N}{4\pi\lambda^2\cos^4\anglepsi}\right)^{2/3}
\Bigl\lbrack
-\rescaledU^{4}\Delta^2 dt^2
+4\pi\lambda\cos^4\anglepsi\,
  \rescaledU^{-2}\Delta^{-4}\sum_i dx_i^2 
+\rescaledU^2\Delta^2 d\dualphi^2
\cr &&
+ 4\pi\lambda\Delta^2\cos^4\anglepsi
\Bigl(d\rescaledU^2  + \rescaledU^2 \dsBase{2}^2 
+\sum_{i=8,9} d\rescaledY_i^2 
\Bigr) 
+N^{-2}\lambda^2
\Delta^2\rescaledU^2 d\Mcircle^2
\Bigr\rbrack
\ .
\label{eqn:Mlift}
\eeq
There is also a nonzero 3-form that we will not write down.
It is interesting to note that as
$\rescaledU\rightarrow\infty$,
the metric (\ref{eqn:Mlift}) becomes more and more flat
(for $\anglepsi\neq\pi/2$), and this suggests that
the ultrahigh energy regime of noncompact PFT
(all $R_i=\infty$) is holographically dual to a weakly coupled
M-theory background.
On the other hand, if $R_i$ is finite
the size of the $x_i$ direction in (\ref{eqn:Mlift})
becomes comparable to the Planck scale for
\be
\rescaledU\sim\frac{\lambda^{-1/6}N^{1/3}R_i}{\Delta^2}\ .
\ee
Beyond that scale, the lift to M-theory is insufficient,
and more complicated duality transformations are in order.
There is in fact an intricate phase structure depending
sensitively on the rationality of $\chi$ 
(or how well it is approximated by a rational number
with a given denominator), which we will study in detail in
section~\ref{sec:Renormalization}.

\end{enumerate}

\section{Thermodynamics of Puff Field Theory}
\label{sec:Thermodynamics}

A simple observable one can compute from the supergravity dual is the entropy. 
The entropy as a function of temperature can be extracted from
the finite temperature
generalization of the dual supergravity solution by
applying the Beckenstein-Hawking formula
\cite{Callan:1996dv,Breckenridge:1996is}.
The finite temperature solution is also easy to construct for PFT.
One simply starts with the smeared non-extremal D0-brane solution instead of (\ref{IIA}) which has the following form:
\beq ds^2 &=& -\fThermal h^{-1/2} dt^2 
 + h^{1/2} \sum_{i=1}^3 dx_i^2
 + h^{1/2} \fThermal^{-1} d \rho^2 \cr
&& \qquad 
+ \rho^2 \bigl(d \anglepsi^2 + \sin^2\anglepsi\,d\angleY^2 
+ \cos^2\anglepsi \lbrack 
\dsBase{2}^2 + (d\fiberphi + \connHopf)^2
\rbrack\bigl) 
\cr
A & = & h^{-1} dt \cr
e^{2 \dilaton} &=& h^{3/2}  \eeq
where
\be \fThermal\equiv 1 - {\rho_0^4 \over \rho^4}\ee  
is the ``thermal factor,''
and \textit{$\dsBase{2}^2$} and $\connHopf$ were defined 
in (\ref{eqn:dsBase}) and (\ref{eqn:defconnHopf}), respectively.
Applying the same set of transformations, we arrive at a solution
\beq {ds^2\over \alpha'} & = & 
\rescaledAugH^{1/2} \left( -\fThermal\rescaledH^{-1} dt^2
+ \fThermal^{-1}d\radialV^2  
+ \radialV^2
\bigl(\cos^2\anglepsi\,\dsBase{2}^2 
+\sin^2\anglepsi\,d\angleY^2 +d\anglepsi^2\bigr) \right) 
\cr && \qquad
+\rescaledAugH^{-1/2} \left(
\sum_i dx_i^2 
+ \rescaledH\radialV^2\cos^2\anglepsi\,
(d \fiberphi + \connHopf + \Delta^3\rescaledH^{-1} dt)^2
\right)\ ,
\cr
{A \over \alpha'^2} & = &  
\rescaledAugH^{-1}
\left( -  dt + \radialV^2 \Delta^3 \cos^2\anglepsi\,
(d \fiberphi + {\cal A}) \right) 
\wedge dx_1 \wedge dx_2 \wedge dx_3\ ,
\cr
e^\dilaton & = & \gIIB = 2 \pi \gYM^2\ , \label{eqn:ThermalSUGRA}
\eeq
where
\be \fThermal = 1 - {\radialV_0^4 \over \radialV^4}\ , 
\ee
and
\be
\rescaledH = \frac{4\pi\gIIB N}{\radialV^4}\ ,
\qquad
\rescaledAugH = 
  \rescaledH + \cos^2\anglepsi\,\Delta^6\radialV^2\ ,
\ee
in accordance with (\ref{eqn:defrescaledHAugH}).
Here \textit{$\radialV_0$} is a free parameter,
and the background (\ref{eqn:ThermalSUGRA}) reduces
to (\ref{eqn:dualSG}) for \textit{$\radialV_0=0$}.
Similar constructions of non-extremal solutions in
asymptotically non-trivial geometries have also appeared in
\cite{Gimon:2003xk}.

The $\radialV=\radialV_0$
hypersurface corresponds to the horizon in this geometry.
In order to extract the thermodynamic behavior of entropy
$S(T)$ as a function of temperature,
it is useful to first determine the temperature $T$ 
and the horizon area $\Area$ as a function of the horizon
radius \textit{$\radialV_0$}.
As usual, the Hawking temperature can be inferred from 
the condition that the Euclidean continuation of this solution
be singularity-free. This gives
\be \radialV_0 = 2\pi\sqrt{\pi\lambda} T\ ,
\label{eqn:UT}
\ee
which is also one of the standard derivations of the 
energy-distance relation (\ref{uvir}).

In order to apply the Beckenstein-Hawking formula, we need
the area of the horizon in the Einstein frame.
In string frame, we need 
\be A = {1 \over \alpha'^4} e^{-2 \dilaton} \sqrt{g_s} \ , \ee
where $g_s$ is the determinant of the induced metric
on the horizon.
This formula, applied to  (\ref{eqn:ThermalSUGRA}), gives 
\be S = {1 \over \gIIB^2} 
\sqrt{ \gIIB N} (2\pi)^3\Volume \radialV_0^3 = 
N^2 (2\pi)^3\Volume T^3 , \label{entropy}\ee
where \textit{$\Volume =  R_1  R_2  R_3$}. This is our main
conclusion concerning the entropy. 
Note that the final expression is
independent of the ``puffness'' $\Delta$. It should be emphasized, however, that this result is reliable only
in the range of temperatures
\be {\lambda^{-1/4} \over  R_i} \ll T \ll {\lambda^{-1/2}  R_i^2  \over \Delta^3} \ , \ee
which does depend on the puffness.  This is similar to what was found in non-commutative gauge theory
\cite{si99,Maldacena:1999mh,Hashimoto:1999yj}. 
We will comment further
on the implication of the range of validity in 
section~\ref{sec:Renormalization}.

The metric (\ref{eqn:ThermalSUGRA}) also contains
information about the chemical potential conjugate
to R-charge.
Like PFT itself,
the metric  preserves an
$SU(2)\times U(1)$ subgroup of the R-symmetry group $SU(4).$
The $U(1)$ component is generated by $\partial/\partial\fiberphi$,
and the chemical potential conjugate to the corresponding
R-charge can easily be read-off from
the solution (\ref{eqn:ThermalSUGRA}).
It is the angular velocity of the event-horizon:
\be
\mu_R \equiv \Omega_{\text{horizon}} = 
\Delta^3 \rescaledH^{-1}\Bigr\rvert_{\text{horizon}} = 
\frac{\eta {\alpha'}^2}{4\pi \gIIB N}\radialV_0^4 
 = 4 \pi^5 \lambda \Delta^3 T^4 \ . 
\ee
The chemical potential depends on the puffness $\Delta$,
and it would be interesting to find the holographic dual
for zero chemical potential, for which
the entropy might also depend on $\Delta.$
We hope to report on this in a separate paper.

\section{Renormalization Group Flow and Hierarchy of PFT}
\label{sec:Renormalization}

The supergravity solution (\ref{eqn:dualSG})
is reliable in the range (\ref{eqn:Urange}).
It is natural to contemplate what alternative
description takes over as a reliable description outside this range.
Precisely such an issue, in the context of 
non-commutative gauge theory,
was investigated in \cite{Hashimoto:1999yj}. 
We will see below that
the PFT case is quite similar.

The infra-red boundary
of the region of validity (\ref{eqn:Urange})
has a simple interpretation: 
at sufficiently low energies the higher
dimensional operators deforming the theory become irrelevant, 
and one simply undergoes dimensional 
reduction below the scale of the size
of the compactification. Let us assume for simplicity that 
\textit{$R_1$}, \textit{$R_2$} and \textit{$R_3$}
are of the same order of magnitude.
{}From the string theory dual point of view,
T-duality along the compact directions maps the D3-branes
to D0-branes.
One expects the Gregory-Laflamme instability to localize the
smeared supergravity solution, simply giving rise to the near horizon
geometry of the D0-branes as the effective description beyond the
region of validity (\ref{eqn:Urange}) on the IR side
\cite{Itzhaki:1998dd,Brandhuber:1998er}.

The proper size of the compact direction also becomes
sub-stringy at the other end of the region of validity
(\ref{eqn:Urange}), i.e., at the upper bound on $\rescaledU.$ 
One does not expect the same three T-dualities to transform this
background to a description which is effective. 
The lesson from non-commutative gauge theories and NCOS
prompts us to look for more complicated U-duality transformations
that can make that region of the background weakly coupled.
In the case of
non-commutative gauge theories and NCOS, 
the appropriate transformations
are elements of $SL(2,Z)$, namely T- and S-duality transformations,
respectively \cite{Hashimoto:1999yj,Chan:2001gs}. 
For PFT, we propose the following $SL(2,Z)$ transformation:
\begin{enumerate}
\item First,  T-dualize along $x_1$, $x_2$, and $x_3$ and lift to M-theory. This brings us back to (\ref{Mtheory}) where $z$ is a periodic coordinate with radius
\be R = \gIIA \lst\ . 
\label{eqn:defR}\ee
\item Now, perform a coordinate transformation
\be \left(\begin{array}{c} d \fiberphi  \\ {d z \over  R} \end{array} \right) 
\rightarrow \left( \begin{array}{cc} \sty{a} & \sty{b} \\ \sty{c} & \sty{d} \end{array}\right)
\left(\begin{array}{c} d \fiberphi  \\ {d z \over  R} \end{array} \right),
\qquad
\left( \begin{array}{cc} \sty{a} & \sty{b} \\ \sty{c} & \sty{d} \end{array}\right)
\in SL(2,Z)\ ,
\ee
which makes the metric take the form
\beq ds^2 &=&  - h^{-1} dt^2  
+   h (\sty{d}\,d  z + \sty{c}  R d \fiberphi - h^{-1} dt)^2  
+\sum_{i=1}^3 dx_i^2 + d r^2 
+ r^2  \dsBase{2}^2 \cr
&&   +  r^2\left( {(  \sty{d} \eta  R  + \sty{b})d  z \over  R} 
+ {(\sty{a} + \sty{c}   \eta  R) d \fiberphi 
+ \connHopf}\right)^2
 +\sum_{i=8}^9 d  y_i^2\ .
\eeq

\item Reduce to IIA along the $z$ direction. There are several subtleties in performing this step.  
At this point it is convenient to set
\be z = {1 \over \sty{d}} \tz\ , \ee
where $\tz$ has the periodicity
$\tz \sim \tz + 2 \pi \sty{d} R$, and
\beq ds^2 &=&  - h^{-1} dt^2  
+   h (d  \tz + c  R d \fiberphi - h^{-1} dt)^2  
+\sum_{i=1}^3 dx_i^2 + d r^2 + r^2  \dsBase{2}^2 \cr
&&   +  r^2\left( {( \eta  R  + {\sty{b} \over \sty{d}} )d  \tz \over  R} + {(\sty{a} + \sty{c}  \eta  R ) d \fiberphi 
+ \connHopf}\right)^2
 +\sum_{i=8}^9 d  y_i^2\ .
\eeq
The IIA solution after the reduction then has the form
\beq ds^2 & = & \sqrt{
h + {(\sty{b}+ \sty{d} \eta R)^2 r^2 \over \sty{d}^2 R^2}} \Bigl( - h^{-1} dt^2+
\sum_i dx_i^2 + dr^2  + r^2\dsBase{2}^2  \cr
&& + 
 \frac{r^2 ( \sty{d} h R \, \connHopf + h R \, d \fiberphi +
  \left( \sty{b} + \sty{d} \eta R \right) dt
)^2}{h \left( \sty{d}^2 h R^2 + r^2(\sty{b} + \sty{d} \eta R)^2 \right) }
 +\sum_{i=8,9} dy_i^2 \Bigr)\ , \cr
A & = &  \sty{c} R d \fiberphi + {R (\sty{b}+\sty{d} \eta R) r^2 (d \fiberphi + \sty{d} \connHopf) - \sty{d}^2 R dt \over \sty{d}^2 R^2 h+ (\sty{b} + \sty{d} \eta R)^2 r^2} \cr
e^\dilaton & = & \tgIIA \left(h + {(\sty{b}+ \sty{d} \eta R)^2 r^2 \over \sty{d}^2 R^2}\right)^{3/4}\ ,
\eeq
for which the string coupling constant and the tension change to
\be \tgIIA = \sty{d}^{3/2} \gIIA, \qquad 
\tilde \alpha' = {1 \over \sty{d}} \alpha'\ , 
\label{eqn:tgta}\ee
because the choice of M-theory circle is different.

\item T-dualize along $x_1$, $x_2$, and $x_3$. This brings the background to the form
\beq ds^2 & = & \sqrt{
h + {(\sty{b}+ \sty{d} \eta R)^2 r^2 \over \sty{d}^2 R^2}} 
\Bigl( - h^{-1} dt^2+
{1 \over
h + {(\sty{b}+ \sty{d} \eta R)^2 r^2 \over \sty{d}^2 R^2}}
\sum_i dx_i^2 + dr^2  + r^2\dsBase{2}^2  \cr
&& + 
 \frac{r^2 ( \sty{d} h R \, \connHopf + h R \, d \fiberphi +
  \left( \sty{b} + \sty{d} \eta R \right) dt
)^2}{h \left( 
\sty{d}^2 h R^2 + r^2(\sty{b} + \sty{d} \eta R)^2
\right) }
 +\sum_{i=8,9} dy_i^2 \Bigr)\ , \cr
A & = &  \left(\sty{c} R d \fiberphi + {R (\sty{b}+\sty{d} \eta R) r^2 (d \fiberphi + \sty{d}\connHopf) - \sty{d}^2 R dt \over \sty{d}^2 R^2 h+ (\sty{b} + \sty{d} \eta R)^2 r^2}\right) \wedge dx_1 \wedge dx_2 \wedge dx_3\ , \cr
e^\dilaton & = & \tgIIB\ .
\eeq
The radii of the $x_i$ coordinates are
\be \tilde R_i = {\tilde \alpha'  \over \alpha'} R_i 
=  {R_i \over \sty{d}}\ .  \ee
We also find 
\be \tgIIB = \tgIIA \tVolume \tilde\alpha'^{-3/2} =  
\gIIB \  , \qquad 
\tVolume = \tilde R_1 \tilde R_2 \tilde R_3\ . 
\label{eqn:deftgIIB}\ee
In terms of $\tilde\alpha'$ and $\tgIIB=\gIIB$,
\be h = 1 + {4 \pi \gIIB N \sty{d}^2 \tilde\alpha'^2 \over r^4}\ ,
\ee
indicating that the number of D3-branes has become $\sty{d}^2 N$. However, we see that near $r=0$, the $d\fiberphi^2$ component of the metric has the form
\be {1 \over \sty{d}^2} r^2 (\sqrt{h}) d \fiberphi^2 \ee
indicating that there is a  $Z_\sty{d}$ orbifold singularity. 
The total D3-brane charge is therefore $\sty{d}N$.

\item Now, we take the decoupling limit sending 
$\tilde \alpha'\rightarrow 0$ keeping 
$\dualizedRescaledU \equiv r/\tilde \alpha'$, 
$\tgIIB = \sty{d} R \tVolume /\tilde \alpha'^2$
[using
(\ref{eqn:defR}),(\ref{eqn:tgta}) and (\ref{eqn:deftgIIB})],
and $\chi = \eta R.$
This brings the SUGRA solution to the form
\beq {ds^2\over \alpha'} & = & 
\dualizedRescaledAugH^{1/2} \left( 
- \dualizedRescaledH^{-1} dt^2
+ d\dualizedRescaledU^2  + \dualizedRescaledU^2 \dsBase{2}^2 
+\sum_{i=8,9} d\rescaledY_i^2 \right) 
\cr && \qquad
+\dualizedRescaledAugH^{-1/2} \left(
\sum_i dx_i^2 
+ \dualizedRescaledH\dualizedRescaledU^2
({d \fiberphi \over \sty{d}}
 + \connHopf + \Delta^3\rescaledH^{-1} dt)^2
\right)\ ,
\cr
{A \over \alpha'^2} & = &  
\left\{ {\sty{c} \over \sty{d}} \tgIIB\tVolume^{-1} d \fiberphi 
+\dualizedRescaledAugH^{-1}
\left( -  dt + \dualizedRescaledU^2 \dualizedDelta^3 
({d \fiberphi \over \sty{d}} + {\cal A}) \right)\right\} 
\wedge dx_1 \wedge dx_2 \wedge dx_3\ ,
\cr
e^\dilaton & = & \tgIIB\ ,  \label{sl2zsol}
\eeq
where
\be \dualizedDelta^3 \equiv 
{(\sty{b}+ \sty{d} \chi) \tVolume \over \tgIIB}\ ,
\qquad
\dualizedRescaledH \equiv
{4 \pi \tgIIB N \over 
(\dualizedRescaledU^2 + \|\rescaledY\|^2)^2}\ ,
\qquad
\dualizedRescaledAugH \equiv 
\dualizedRescaledH + \dualizedRescaledU^2\dualizedDelta^6\ .
\ee
\end{enumerate}
This is the SUGRA dual of the $SL(2,Z)$ transform of PFT.  It has the same form as the SUGRA dual of PFT (\ref{eqn:dualSG}),
except that the $\fiberphi$ coordinate has a deficit angle, 
and there is an extra constant term in the RR 4-form potential
(which cannot be gauged away because the $x_1, x_2, x_3$
and $\fiberphi$ directions are compact).

The $SL(2,Z)$ transformation also acts non-trivially on the D3-brane charge, the volume of the torus, and the puffness. Specifically
\be N \rightarrow \tilde N = \sty{d}^2 N, \qquad 
R_i \rightarrow
\tilde R_i = R_i/\sty{d}, \qquad 
\chi \rightarrow \tilde \chi =
\sty{b} + \sty{d} \chi\ , \ee
which is the analogue of the Morita transformation formula 
of NCYM \cite{Hashimoto:1999yj,Pioline:1999xg}.
Note that the UV/IR relation
\be E = {\dualizedRescaledU \over \sqrt{\tilde \lambda}} 
   = {\rescaledU \over \sqrt{\lambda}} \ee
gives rise to a consistent holographic embedding. The constant part of
the RR 4-form in PFT is the analogue of the ``$\Phi$ parameter'' 
in NCYM.

We can immediately infer the range of validity of this solution 
\be {\tilde \lambda^{-1/4} \over  \tilde R_i} \ll E \ll {\tilde \lambda^{-1/2}  \tilde R_i^2  \over \dualizedDelta^3}  = 
{\tgIIB \tilde \lambda^{-1/2} \over  \tilde\chi} 
{1 \over \tilde R_i}  \ . \label{Erange2} \ee
Computing the entropy from the near extremal generalization of (\ref{sl2zsol})
yields
\be S = {1 \over \sty{d}} (\sty{d}^2 N)^2 (2\pi)^3\tVolume T^3 
= (2\pi)^3 N^2\Volume T^3 \ . \ee
In other words, the functional form of the entropy formula
(\ref{entropy}) appears to extend beyond its naive range of
applicability, as long as there is some dual description in terms 
of one of the $SL(2,Z)$ duals listed in (\ref{sl2zsol}).

If the value of $\chi$ of PFT we start with is rational, say,
\be \chi = {\sty{r} \over \sty{s}} \ee
then there exists an element of $SL(2,Z),$
\be \left(\begin{array}{cc}\sty{a} & \sty{b} \\\sty{c} & \sty{d}\end{array}\right) = 
 \left(\begin{array}{cc}\sty{p} & -\sty{r} \\-\sty{q} & \sty{s}\end{array}\right)\ ,  \ee
for which $\dualizedDelta = 0,$ and the solution becomes
\beq {ds^2\over \tilde \alpha'} & = & 
- \dualizedRescaledH^{-1/2} dt^2
+ \dualizedRescaledH^{-1/2} \sum_i dx_i^2 
+ \dualizedRescaledH^{1/2} \left( d\dualizedRescaledU^2 
+ \dualizedRescaledU^2 \dsBase{2}^2  + 
\dualizedRescaledU^2
\left({d \fiberphi \over \sty{s}}+  \connHopf \right)^2\right)
 \cr
{A \over \tilde \alpha'^2}  & = &  \left( -{ \sty{q} \over \sty{s}} \tgIIB\tVolume^{-1} d \fiberphi 
- \dualizedRescaledH^{-1}  dt \right)
   \wedge dx_1 \wedge dx_2 \wedge dx_3 \cr
e^\dilaton & = & \tgIIB  \label{localsg}
\eeq
which is essentially $AdS_5 \times S_5 / Z_\sty{s}$ with constant RR 4-form potential.
The fact that this solution is  anti de-Sitter 
provides further evidence that gravity is decoupled from the
dynamics of PFT.

\begin{figure}
\centerline{\includegraphics[width=4in]{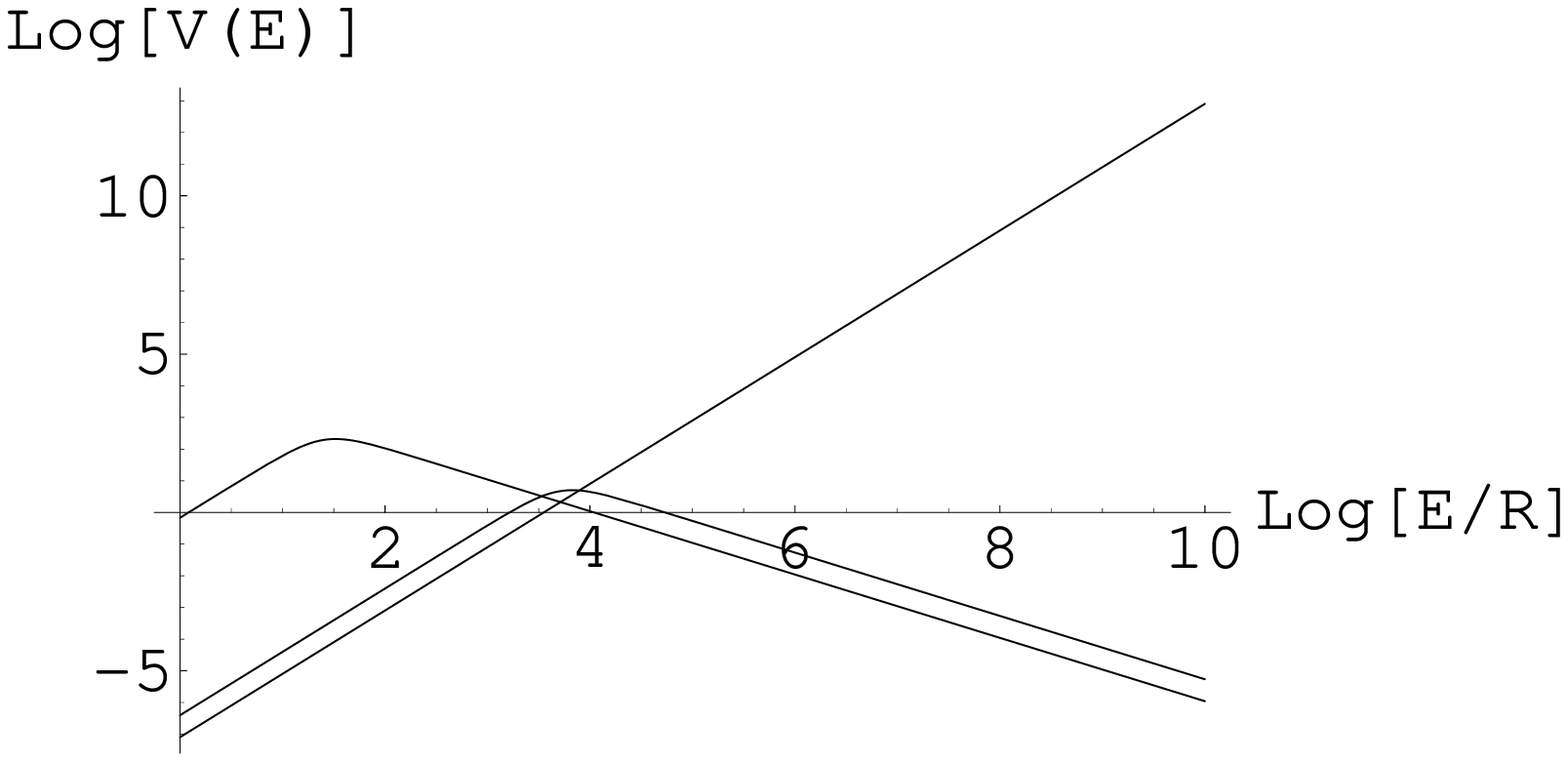}}
\caption{Log-log plot of  (\ref{volume})  for $(\sty{b},\sty{d}) = (0,1)$, $(\sty{b},\sty{d}) = (-1,511)$, and $(\sty{b},\sty{d}) =
(-2,1023)$ for a PFT with parameters $2 \pi \gYM^2 = \gIIB = 1/3$, $\lambda = 9$, and $\chi =
2/1023$.
\label{figa}}
\end{figure}

The region of validity of (\ref{localsg}) is
\be {\tilde \lambda^{-1/4} \over \tilde R} < E  
\label{Erange3} \ . \ee
Provided $\chi$ is chosen such that
\be 1 \ll {\sqrt{\sty{s}} \lambda^{1/4} \over \gIIB \chi}\ , \ee
which is easy to arrange since we assumed $\lambda \gg 1$ and 
$\gIIB \ll 1$,
 the upper bound of (\ref{Erange2}) is smaller than the lower bound of (\ref{Erange3}):
\be {\lambda^{-1/2} R^2 \over \Delta^3} \ll {\tilde \lambda^{-1/4} \over \tilde R}\ . \ee
As the supergravity dual (\ref{localsg}) does not have an upper bound on its region of applicability (\ref{Erange2}), 
one can conclude that any PFT with rational value of $\chi$ 
is described in terms of it.
Since rational $\chi$'s form a dense subset of the set of real values of $\chi$, 
we conclude that for arbitrary values of $\chi$, the entropy
formula (\ref{entropy}) is valid for all energies,
\be {\lambda^{-1/4} \over R_i} < E \ , \label{Erange4}\ee
assuming that the entropy is a continuous function of $\chi.$
The specific $SL(2,Z)$ element which gives the most effective
description at a given energy $E$ in the range (\ref{Erange4}) does,
however, depend sensitively on the rationality of $\chi$.  
To determine which $SL(2,Z)$ is most effective, one looks for 
a pair $(\sty{b},\sty{d})$ that maximizes the proper size
of the $x_i$ circle, or equivalently the expression
\be \VExpression(E) \equiv
{\tilde R^4 \over \dualizedRescaledH^2 
  + \dualizedDelta^6\dualizedRescaledU^2} 
=
{\gIIB^2  R^2  \lambda^2 E^4 \over 4 \pi \sty{d}^2 \gIIB^2 
 +  (\sty{b}+ \sty{d}\chi)^2  \lambda^2 R^6 E^6} \ . 
\label{volume} 
\ee
For example,
take $\gIIB = 1/3$, $\lambda = 9$, and $\chi = 2/1023.$
We find that $(\sty{b},\sty{d}) = (0,1)$, 
$(\sty{b},\sty{d}) = (-1,511)$, 
and $(\sty{b},\sty{d}) = (-2,1023)$ give rise to a 
$\VExpression(E)$ that is illustrated in
figure \ref{figa}. 
Similar structures were encountered in the case of
non-commutative gauge theory \cite{Hashimoto:1999yj} and NCOS
\cite{Chan:2001gs}, where a self-similar structure,
closely related to
the continued fraction expansion for the appropriate counterparts 
of the dimensionless non-locality parameter $\chi$,
characterizes the phase diagram.

Here, the fact that the range of validity (\ref{Erange2}) depends on
two large dimensionless parameters $\tilde\lambda$ and $1/\tgIIB$, in addition to $\chi$, makes the full phase structure
somewhat more cumbersome to determine.  For example, 
for $\dualizedDelta\ll\tgIIB^{1/3} {\tilde\lambda}^{-1/6} R$
[cf. (\ref{eqn:ConditionDelta}),
which is stronger than (\ref{eqn:assump})
for $\lambda^{1/4}\gg\gIIB$],
we need to go to the M-theory description as we explained 
at the end of section~\ref{sec:SupergravityDual}.
This description
does not drastically alter the form of the entropy 
as a function of temperature, because
the entropy formula (\ref{entropy}),
which is based on the area of the horizon 
in Einstein frame, is generally 
unaffected by T-dualities and by the M-theory lift.
Strictly speaking, we have not ruled out the possibility
of some exotic thermodynamic behavior in the range of energies for
which the supergravity description is not effective, along the lines
of what was observed in \cite{Chan:2001gs}. Nonetheless, one expects
some specific dual description to be effective for any range of
parameters and energies. In these duality cascades, the fact that
there is a PFT in the far IR and (\ref{localsg}) in the far UV
for any rational value of $\chi$ appears to be a robust feature.

This also highlights the point that the decompactification 
limit $R_i\rightarrow \infty$ is a tricky limit to take even if one concentrates on the UV.
This is because making $R_i$ large while keeping $\Delta$ fixed
changes the rationality of $\chi$ in a chaotic way. 
While physical observables, such as the entropy, 
have a smooth limit, the phase
structure in the UV region evolves erratically. Such interference
between flows to the UV and decompactification is a typical feature of non-local field theories 
\cite{Hashimoto:1999yj,Chan:2001gs}.

\section{Deformation Operator of Lowest Dimension}
\label{sec:Deformation}

In the limit $\rescaledU\rightarrow 0$
the supergravity dual (\ref{eqn:dualSG}) becomes 
$AdS_5\times S^5$,
which corresponds to PFT flowing to \SUSY{4} SYM in the IR.
The supergravity dual (\ref{eqn:dualSG})
can also be used to read off the
lowest dimension operator responsible 
for deforming the \SUSY{4} theory. 
We see in (\ref{eqn:dualSG}) that a linear combination of the
metric and the RR 4-form potential,  polarized partly along the brane and partly transverse to the brane, are deformed. 
The deformation of $AdS_5\times S^5$
that (\ref{eqn:dualSG}) describes
has been arranged to preserve half of the supersymmetries, 
and therefore the corresponding operator
has to be a descendant of a chiral primary operator.

We denote the \SUSY{4} SYM gauge field strength by
$F_{\mu\nu}$, the scalars by $X^I$ ($I=4,\dots,9$ for convenience),
and the spinors by $\lambda$ and $\bar{\lambda}$.
(We will not need to specify the indices on the spinors.)
The descendents of chiral primary operators of \SUSY{4} SYM
are listed in table~7 of \cite{D'Hoker:2002aw}.
In their notation,
our requisite descendant takes the form
\be 
\Op{17}{k} \sim
\tr F_+ F_- \lambda \bar \lambda X^k \ee
for $k=0.$
The schematic notation here is as follows:
$F_{+}$ ($F_{-}$) stands for the self-dual (anti-self-dual)
part of the field-strength, $X^k$ stands for
a product of $k$ scalar fields, there is
an unspecified index contraction, and terms involving derivatives
and commutators have been suppressed.
$\Op{17}{k=0}$
is an operator of dimension 7 in the representation
${\bf 15}$ of the
$SO(6)$ $R$-symmetry group. This $SO(6)$ multiplet accounts for
distinct ways in which the space $\Reals^4$, which we twist, can be
embedded into the $\Reals^6$ space transverse to the D3-brane.

Let us note, in contrast, that the leading irrelevant operators
that deform \SUSY{4} SYM into SYM on a noncommutative
$\Reals^4$ (NCYM)
and the noncommutative open string theory (NCOS) are,
respectively,  
the real and the imaginary parts of the dimension 6 operator 
\be 
\Op{16}{k=0}\sim
\tr F_+ F_-^2 X^{k=0}\ ,  \ee
whereas the dipole deformation and its S-dual
are generated by the real and imaginary parts of the
dimension 5 operator
\be
\Op{10}{k=0}\sim
\tr F_+ \lambda \bar \lambda X^{k=0} \ . \ee
These dimensions fit well with the fact that the parameters
characterizing the dipole, the non-commutative, and the 
puff field theories have dimensions 1, 2, and 3, respectively.

PFT can also be defined as the decoupled field-theory
that describes $N$ D3-brane probes in
the strongly-coupled type-IIB background
obtained from (\ref{eqn:FullSolution}) by setting $N=0.$
This is the Melvin background that can be written as
\beq ds^2 & = & 
(1+\eta^2\fullu^2)^{1/2} \left( - dt^2
+ d\fullu^2  + \fullu^2 \dsBase{2}^2 
+\sum_{i=8,9} dy_i^2 \right) 
\cr && \qquad
+(1+\eta^2\fullu^2)^{-1/2} \left(
\sum_i dx_i^2 
+ \fullu^2(d \fiberphi + \connHopf + \eta dt)^2
\right)\ ,
\cr
A & = &  \frac{1}{1+\eta^2\fullu^2}
\left( -  dt + \fullu^2 \eta (d \fiberphi+\connHopf) \right) 
\wedge dx_1 \wedge dx_2 \wedge dx_3\ ,
\cr
e^\dilaton & = & \gIIB\ .
\label{eqn:FullSolutionZ}
\eeq
It is strongly coupled in the limit $\alpha'\rightarrow 0$
keeping (\ref{eqn:defDelta}).
The operator $\Op{17}{k=0}$ can be interpreted as follows.
Expand (\ref{eqn:FullSolutionZ}) formally in powers of $\eta$,
and keep only terms up to order $O(\eta)$.
Using the notation
$$
\angularForm\equiv
\fullu^2(d\fiberphi + \connHopf) 
= y_4 dy_5 - y_5 dy_4 + y_6 dy_7 - y_7 dy_6\ ,
$$
we can write (\ref{eqn:FullSolutionZ}) as
\beq ds^2 & = & 
-dt^2 + \sum_i dx_i^2 
+ \sum_{i=4}^9 dy_i^2+2\eta\omega dt + O(\eta^2)\ ,
\cr
A_4^{\text{\tiny (full)}} & = &
(-dt +\eta\omega) 
\wedge dx_1 \wedge dx_2 \wedge dx_3
+\eta\omega\wedge dt\wedge dy_8\wedge dy_9 + O(\eta^2)\ ,
\cr
e^\dilaton & = & \gIIB\ ,
\label{eqn:SpecialMelvinOrderEta}
\eeq
where we have completed the RR 4-form so that
$dA_4^{\text{\tiny (full)}}$ is self-dual.
The bosonic part of $\Op{17}{k=0}$ can now be deduced
from the Dirac-Born-Infeld (DBI) action and
the Wess-Zumino (WZ) term,
$$
{\cal S}_{\text{\tiny DBI+WZ}} = 
\frac{1}{{\alpha'}^2\gIIB}\int_{D3}\left(
\sqrt{-\det{G + \alpha' F}}\,
+A_4^{\text{\tiny (full)}}\right)\ ,
$$
where the induced metric $G$ is given by
$$
G_{\mu\nu} = \eta_{\mu\nu} 
+{\alpha'}^2 \sum_{i=4}^9 \partial_\mu X^i\partial_\nu X^i
+\eta {\alpha'}^2 (\delta_{\mu 0} \RCurrent_\nu 
+ \delta_{\nu 0}\RCurrent_\mu) + O(\eta^2)\ ,
$$
$\RCurrent_\mu$ is the R-current:
$$
\RCurrent_\mu\equiv
 X^4\partial_\mu X^5 - X^5\partial_\mu X^4
+X^6\partial_\mu X^7 - X^7\partial_\mu X^6\ ,
$$
and we used the standard relation 
$y_i = \alpha' X^i$ ($i=4\dots 9$) between
the transverse coordinates of the D3-brane and the
scalar fields of the effective field theory on the brane.

Expanding the DBI action to order $O(\eta)$ we find, for $N=1$,
\beq
\Op{17}{k=0} &=& T^{0\mu}\RCurrent_\mu
+\epsilon^{0\mu\nu\sigma}
\partial_\mu X^8\partial_\nu X^9\RCurrent_\sigma
+\text{fermions}\ ,
\label{eqn:Op17}
\eeq
where
$$
T^{\mu\nu} = \sum_{i=4}^9\partial^\mu X^i\partial^\nu X^i
-\frac{1}{2}\eta^{\mu\nu}\sum_{i=4}^9
  \partial_\tau X^i\partial^\tau X^i
+{F^\mu}_{\tau}F^{\tau\nu}
+\frac{1}{4}\eta^{\mu\nu}F_{\sigma\tau}F^{\sigma\tau}
+\text{fermions}\ ,
$$
is the stress-energy tensor.
For $N>1$, (\ref{eqn:Op17}) is missing an overall trace
and additional commutator terms.

\section{Concluding remarks}
\label{sec:Concluding}

In this paper we inferred a number of basic features 
of Puff Field Theory by analyzing its supergravity dual.
In particular, we computed
the thermodynamic entropy, studied its range of validity, and
identified the leading irrelevant operator deforming the \SUSY{4}
theory.
These results lend more credence to the conjecture that PFT
is decoupled for gravity. In fact, the mere existence of
a (geodesically complete) near-horizon limit
of the background (\ref{eqn:FullSolution}) implies decoupling.
The finite entropy (\ref{entropy}) suggests that the
spectrum is discrete (for appropriate boundary conditions
that eliminate the zero modes of the low-energy scalar fields).
Furthermore, we have seen that for rational $\chi$
the supergravity dual can be transformed into
an orbifold of $AdS_5\times S^5$ 
with extra RR flux (\ref{localsg}), which certainly
describes a decoupled theory.

It would, of course, be interesting if a microscopic definition 
of PFT can be found. 
Non-commutative Yang-Mills theory and dipole theories
can be formulated in terms of a concrete action, and NCOS can be
defined as a strong coupling limit of NCYM. It would be nice if PFT
can be defined at the same level of specificity.

Lessons from NCYM and dipole theories suggest that a good 
starting point might be to study
PFT on $T^3$ with a rational parameter $\chi.$
One approach might be to identify the field theory dual of
(\ref{localsg}). This rather innocent looking supergravity solution
contains a closed RR 4-form potential which, combined with the
orbifold, is responsible for all the non-trivial IR physics. 
We are currently investigating this issue and 
we hope to report our findings in the near future.

PFT arose as the decoupling limit of D0-branes in a Melvin universe
supported by an RR 1-form potential in the type IIA theory.
It is also natural to consider what happens for
other type-IIA D$p$-branes.

For the case of D2-branes, there are two possible choices of embedding:
the twisted $\fiberphi$ coordinate could either
be along or transverse to the D2-brane.
If it is along the brane, one ends up with an NCOS, which is
S-dual to the non-commutative gauge theory of
\cite{Hashimoto:2004pb,Hashimoto:2005hy},
dimensionally reduced to 2+1 dimensions.
The supergravity dual of the NCOS (prior to the
dimensional reduction) was discussed in 
\cite{Huang:2005mi,Cai:2006td}.
If the $\fiberphi$ direction is transverse to the brane,
we end up with the S-dual of dipole theory.
In both of these constructions, we are
dimensionally reducing along the non-local direction from the
NCYM/dipole point of view, but the non-locality of the S-dual
survives dimensional reduction.

The case of D4-branes does not appear to have any interesting non-local field theory in the decoupling limit, because
when  D4-branes are lifted to M-theory they 
are extended along the M-theory circle.
The case of NS5-branes appears to lead to a non-local 
deformation of little string theory (LST),
and neither D6-branes nor D8-branes support 
any decoupled field theory, so we will not pursue them further.

We will elaborate on the details of the twisted decoupling of type
IIA D2, D4, and NS5-branes in appendix \ref{appb}. As NCOS arising
from 2+1 was already known, and LST arising from NS5 is already a
non-local theory, the PFT based on D0-branes appears to be rather
special in giving rise to a novel non-local deformation of a local
field theory. 

\section*{Acknowledgements}
We would like to thank
Lisa~Dyson, 
Ian~Ellwood,
Eric~Gimon, 
Thomas~Grimm,
Petr~Ho\v{r}ava,
Christopher~Hull,
Nissan~Itzhaki,
Yasunori~Nomura
and Jesse~Thaler
for discussions.
This work was supported in part by
the Director, Office of Science, Office of High Energy and Nuclear
Physics, of the U.S. Department of Energy under
Contract DE-AC03-76SF00098 and under Contract
DE-FG02-95ER40896, in part by
the NSF under grant PHY-0098840,
and in part by
the Center of Theoretical Physics at UC Berkeley.
OJG also wishes to thank the organizers
of the conference ``M-theory in the City,'' which took place
at Queen Mary University of London in November 2006,
and AH thanks UC Berkeley, where this work was initiated,
for their warm hospitality.

\appendix

\section*{Appendix}

\section{Melvin twists of D2, D4, and NS5 branes\label{appb}}

In this article, we primarily focused on the decoupled 
field theory on D0-branes embedded in a Kaluza-Klein 
Melvin universe with the M-theory
circle playing the role of the Kaluza-Klein circle.  Such a
construction naturally extends to other branes 
in type-IIA string theory.
In this appendix, we elaborate on the cases of D2, D4, and NS5
branes. In all of these cases, the appropriate scaling of the 
Melvin flux can be inferred from requiring the 
dimensionless parameter $\chi$ to be finite.

\subsection{D2-brane}

The decoupled theory on D2-branes turns out to be a known non-local
field theory. In order to identify this field theory, let us analyze
the supergravity dual explicitly.

Let us follow the construction of 
section~\ref{sec:SupergravityDual}.
Start with the
supergravity solution of D2
\beq  ds^2 & = & h^{-1/2} (-dt^2 + \sum_{i=1}^2 dx_i^2) 
+  h^{1/2} \sum_{i=3}^9 dx_i^2 \ ,\cr
A & = & h^{-1} dt \wedge dx_1 \wedge dx_2\ , \cr
e^{\dilaton} & = & \gIIA h^{1/4}\ , \cr
h & = & 1 + {6 \pi^2 \gIIA N \alpha'^{5/2}\over r^5} \ . \eeq
Lifting  to M-theory gives
\beq  ds^2 & = & h^{-2/3} (-dt^2 + \sum_{i=1}^2 dx_i^2) 
+  h^{1/3} \sum_{i=3}^9 dx_i^2 + h^{1/3} dz^2\ ,\cr
A & = & h^{-1} dt \wedge dx_1 \wedge dx_2\ ,  \cr
z & \sim & z + 2 \pi \gIIA \lst \ . \eeq
Twisting along the $(x_1,x_2)$ plane gives
\beq  ds^2 & = & h^{-2/3} (-dt^2 + d\rho^2 
+ \rho^2 (d \fiberphi+ \eta dz)^2) 
+  h^{1/3} \sum_{i=3}^9 dx_i^2 + h^{1/3} dz^2\ ,\cr
A & = & h^{-1} \rho\,  dt \wedge d\rho \wedge 
( d \fiberphi  + \eta dz)\ , \cr
z & \sim & z + 2 \pi \gIIA \lst \ . \eeq
Now, reduce to IIA on $z$ to find
\beq  ds^2 & = & 
\left(h + \eta^2 \rho^2 \over h\right)^{1/2} 
\left\lbrack
h^{-1/2} \left(-dt^2 + d\rho^2 
+ {h \rho^2 \over h + \eta^2 \rho^2} d \fiberphi^2\right) 
+  h^{1/2} (dr^2 + r^2 d \Omega_6^2)\right\rbrack\ ,\cr
A_1 & = & {\eta \rho^2 \over h + \eta^2 \rho^2} d \fiberphi\ , \cr
A_3 & = & h^{-1} dt \wedge dx_1 \wedge dx_2\ ,  \cr
B_2 & = & \eta h^{-1} r\,  dt \wedge dr\ , \cr
e^{\dilaton} & = &  
\gIIA h^{1/4}\left({h + \eta^2 \rho^2 \over h} \right)^{3/4} \ .
\eeq
Finally, taking the $\alpha' \rightarrow 0$ decoupling limit, keeping $U = r/\alpha'$, $\gYMDim{2}^2 = \gIIA \lst^{-1}$
(the YM coupling constant of the 2+1D theory) 
and $\chi = \eta R$ fixed, gives
\beq  {ds^2 \over \alpha'} & = & 
\left(1 + {\chi^2 \rho^2 \over \gYMDim{2}^4 H}\right)^{1/2}
\left\lbrack H^{-1/2} 
\left(-dt^2 + d\rho^2 + {\rho^2 d \fiberphi^2 \over 
1 + {\chi^2 \rho^2 \over \gYMDim{2}^4 H} }\right) 
+  H^{1/2} (dU^2 + U^2 d \Omega_6^2)\right\rbrack\ ,\cr
{A_1 \over \alpha'}  & = & 
{\chi \gYMDim{2}^2 \rho^2  \over 
\gYMDim{2}^4 H + \chi^2 \rho^2} d \fiberphi\ ,\cr
{A_3 \over \alpha'^2} & = & H^{-1} dt \wedge dx_1 \wedge dx_2\ ,
\cr
{B_2 \over \alpha'}  & = & 
{1 \over \gYMDim{2}^2}\chi H^{-1} \rho\,  dt \wedge d\rho\ , \cr
e^{\dilaton} & = &  \gYMDim{2}^2  H^{1/4} 
\left({\gYMDim{2}^4 H 
+ \chi^2 \rho^2 \over \gYMDim{2}^4 H} \right)^{3/4}\ , \cr
H & = & {6 \pi^2 \gYMDim{2}^2 N \over U^5} \label{a5} \ .
\eeq
This is a non-local deformation of a strongly coupled SYM with 16
supercharges in 2+1 dimensions.

In order to bring this theory into context, it is useful to compactify $x_3$ on a circle of radius $\alpha'/R_3$ and smear, so that (\ref{a5}) becomes
\beq  {ds^2 \over \alpha'} & = & 
\left(1 + {\chi^2 \rho^2 \over \gYMDim{2}^4 H}\right)^{1/2}
\left\lbrack
H^{-1/2} \left(-dt^2 + d\rho^2 
+ {\rho^2 d \fiberphi^2 \over 1 
+ {\chi^2 \rho^2 \over \gYMDim{2}^4 H} }\right)\right. 
\cr &&  \qquad\qquad\qquad\qquad\left.
\vphantom{\left({\rho^2 d \fiberphi^2 \over 1 %
+ {\chi^2 \rho^2 \over \gYMDim{2}^4 H}}\right)} %
%
+ {H^{1/2} \over \alpha'^2}dx_3^2   
+ H^{1/2} (dU^2 + U^2 d \Omega_5^2)\right\rbrack\ ,\cr
{A_1 \over \alpha'}  & = & 
{\chi  \rho^2  \over \gYMDim{2}^2 H
(1 + {\chi^2 \rho^2 \over \gYMDim{2}^4 H})} d \fiberphi
\ , \cr
{A_3 \over \alpha'^2} & = & H^{-1} dt \wedge dx_1 \wedge dx_2
\ ,  \cr
{B_2 \over \alpha'}  & = & 
{1 \over \gYMDim{2}^2}\chi H^{-1} \rho\,  dt \wedge d\rho
\ , \cr
e^{\dilaton} & = &  \gYMDim{2}^2  H^{1/4} 
\left({\gYMDim{2}^4 H + \chi^2 \rho^2 \over 
\gYMDim{2}^4 H} \right)^{3/4}\ , \cr
H &\equiv& {8 \pi^2 \gYMDim{2}^2 R N \over U^4}  \ .
\eeq
T-dualizing along $x_3$ brings this to the form
\beq  {ds^2 \over \alpha'} & = & 
\left(1 + {(2 \pi R_3 \chi) \rho^2 \over 
\gYMDim{3}^4 H}\right)^{1/2}
\left\lbrack
H^{-1/2} \left(-dt^2 + d\rho^2 
+ {\rho^2 d \fiberphi^2 + dx_3^2 \over 
1 + {(2 \pi R_3\chi \rho)^2 \over \gYMDim{3}^4 H} }\right) 
\right.
\cr &&\qquad\qquad\qquad\qquad\qquad\qquad\qquad\qquad\qquad
\left.
\vphantom{
\left(-dt^2 + d\rho^2 %
+ {\rho^2 d \fiberphi^2 + dx_3^2 \over %
1 + {(2 \pi R_3\chi \rho)^2 \over \gYMDim{3}^4 H} }\right) %
} %
%
+   H^{1/2} (dU^2 + U^2 d \Omega_5^2)
\right\rbrack\ ,\cr
{A_2 \over \alpha'}  & = & 
{2 \pi R_3 \chi  \rho^2  \over \gYMDim{3}^2 H 
(1 + {(2 \pi R_3 \chi)^2 \rho^2 \over \gYMDim{3}^4 H})} 
d \fiberphi \wedge dx_3\ , \cr
{A_4 \over \alpha'^2} & = & 
H^{-1} dt \wedge dx_1 \wedge dx_2  \wedge dx_3\ , \cr
{B_2 \over \alpha'}  & = & 
{1 \over \gYMDim{3}^2}(2 \pi R_3 \chi) 
H^{-1} \rho\,  dt \wedge d\rho\ , \cr
e^{\dilaton} & = &  \gYMDim{3}^2  
\left(1+{(2 \pi R_3 \chi)^2 \rho^2 \over 
\gYMDim{3}^4 H} \right)^{1/2}\ , \cr
H & = & {4 \pi \gYMDim{3}^2  N \over U^4}\ ,  \cr
\gYMDim{3}^2 & = & 2 \pi R_3 \gYMDim{2}^2 \ .
\eeq
(Here $A_4$ is not the complete RR 4-form,
but is such that the 5-form RR field-strength is the
self-dual part of $dA_4$.)
This solution is the S-dual of the solution of
\cite{Hashimoto:2004pb,Hashimoto:2005hy},
which was also discussed in \cite{Huang:2005mi,Cai:2006td}.
 In other words, (\ref{a5}) can be viewed as a
dimensional reduction of NCOS from 3+1 to 2+1 dimensions.

Twisting instead along a direction transverse to the 
D2-brane gives rise to the S-dual of dipole theories 
\cite{Bergman:2000cw,Bergman:2001rw},
dimensionally reduced from 3+1 to 2+1 dimensions along similar lines.

\subsection{D4-brane}

A similar construction applied to the case of a D4-brane turns out 
{\it not} to give rise to any interesting non-local field theory. 
One reason for this is the fact that a D4-brane, 
when lifted to M-theory,
wraps the M-theory circle unlike the D0 and the D2-branes.  
Also, the radius of the M-theory circle remains finite in the decoupling limit, as can be seen from the relation
\be R = \gIIA \alpha'^{1/2} = \gYMDim{4}^2 = \text{finite.} \ee
The decoupled theory turns out to be nothing more than a twisted
compactification of the decoupled M5 superconformal field theory.

To see this more explicitly, start with the supergravity solution of the 
D4-brane\footnote{We abbreviate the RR 2-form potential as it plays no significant role here.}
\beq  ds^2 & = & h^{-1/2} (-dt^2 + \sum_{i=1}^4 dx_i^2) 
+  h^{1/2} \sum_{i=5}^9 dx_i^2\ , \cr
e^{\dilaton} & = & \gIIA h^{-1/4}\ , \cr
h & = & 1 + {g N \alpha'^{3/2}\over r^3} \ . \eeq
This solution lifts to M-theory as follows
\be  ds^2  =  h^{-1/3} (-dt^2 + \sum_{i=1}^4 dx_i^2 + dz^2) 
+  h^{2/3} \sum_{i=5}^9 dx_i^2 \ . \ee
Now, twist along the world volume,
\be  ds^2  =   h^{-1/3} \left(-dt^2 +   d\rho^2 
+ \rho^2 \left\{
\dsBase{2}^2 +  (d \fiberphi + \connHopf+\eta dz)^2 \right\}
+ dz^2\right) +  h^{2/3} \sum_{i=5}^9 dx_i^2 \ , \ee
and reduce to IIA,
\be  ds^2  =  {1 \over \sqrt{1 + \eta^2 r^2}} 
\left(h^{-1/2} \left\lbrack-dt^2 + d\rho^2 
+ \rho^2 \left\{\dsBase{2}^2 + {1 \over 1+\eta^2 \rho^2} 
(d \fiberphi + \connHopf)^2 \right\}\right\rbrack
 +  h^{1/2} \sum_{i=5}^9 dx_i^2\right) \ . \ee
In the decoupling limit where $R = \gIIA \lst = g_{YM4}^2$ is kept
fixed, $\eta$ also stays finite. This leads to a relatively boring
field theory which is nothing more than 4+1 SYM in a Kaluza-Klein
Melvin universe, which lifts to a M5 SCFT on flat 5+1 dimensional
space-time with twisted compactification
\be ds^2 = -dt^2 +   d\rho^2 + \rho^2 \left\{\dsBase{2}^2 
+  (d \fiberphi + \connHopf+\eta dz)^2\right\}
+ dz^2, \qquad z \sim z + 2 \pi R \ . \ee

\subsection{NS5-brane}

The Kaluza-Klein Melvin background also gives rise to a UV deformation
for the decoupled theory on NS5-branes. To see this, start with the
supergravity solution for NS5-branes\footnote{We ignore the
NSNS 2-form potential here as well.}
\beq
ds^2 & = & -dt^2 + \|d\vec x\|^2
+ h(r) \left(dr^2 + r^2\left\{\dsBase{2}^2 
+ (d \fiberphi + \connHopf)^2\right\}\right), \cr
e^\dilaton & = & g_s h(r)^{1/2}, \cr
h(r) &=& 1 + {m \alpha' \over r^2}\ . \label{ns5bg}
\eeq
Lifting to M-theory, we find
\be 
ds^2  =  h^{-1/3} (-dt^2 + \|d\vec x\|^2) 
+ h^{2/3}(r) \left(dr^2 + r^2\left\{\dsBase{2}^2 
+ (d \fiberphi + \connHopf)^2 + dz^2\right\}\right)\ ,\ee
with $z \sim z + 2 \pi R$.  Twisting gives
\be 
ds^2  =  -h^{-1/3} (-dt^2 + \|d\vec x\|^2) 
+ h^{2/3}(r) \left(dr^2 + r^2\left\{\dsBase{2}^2 
+ (d \fiberphi + \connHopf+ \eta dz)^2 + dz^2\right\}\right)\ . \ee
Reducing to IIA then gives
\be ds^2  =  \sqrt{1 + \eta^2 r^2} \left[ -dt^2 + \|d\vec x\|^2
 + h(r) \left(dr^2 + r^2\left\{\dsBase{2}^2 + 
{(d \fiberphi + \connHopf)^2\over 1+\eta^2 r^2} 
\right\}\right)\right] \ . \ee 

In terms of $\chi = \eta \gIIA \lst$, which we are keeping finite, 
\be ds^2  =  \sqrt{1 + \frac{\chi^2 r^2}{\gIIA^2 \alpha'}} 
\left[ -dt^2 + \|d\vec x\|^2 
+ \left(1 + { N \alpha' \over r^2}\right) 
\left(dr^2 + r^2\left\{\dsBase{2}^2 + {(d \fiberphi + \connHopf)^2\over 1+\chi^2 r^2/\gIIA^2 \alpha'} 
\right\}\right)\right] \ . \ee 
If we now let $r = \gIIA\rho$, and send $\gIIA\rightarrow 0$,
keeping $\alpha'$ fixed,
to derive the dual of the decoupling limit
\be ds^2  =  \sqrt{1 + \chi^2 \rho^2 \alpha'} 
\left[ -dt^2 + \|d\vec x\|^2 
+ { N \alpha' \over \rho^2} 
\left(d\rho^2 + \rho^2\left\{\dsBase{2}^2 
+ {(d \fiberphi + \connHopf)^2\over 1+\chi^2 \rho^2 / \alpha'} 
\right\}\right)\right]\ . \ee 
This is a UV deformation of little string theory in type IIA with
string tension $\alpha'$.  This solution reduces to the
Callan-Harvey-Strominger solution \cite{Callan:1991ky} in the limit
$\chi \rightarrow 0$. However, if the deformation parameter $\chi$ is rational, there exists an $SL(2,Z)$ 
transformation along the lines 
of what is described in section~\ref{sec:Renormalization}
that brings the supergravity solution to the form of a discrete
orbifold of CHS with a constant RR 1-form potential.

\bibliography{PFT}\bibliographystyle{utphys}
\end{document}